\documentclass[aps,prl,twocolumn,showkeys,nofootinbib,preprintnumbers,superscriptaddress,floatfix]{revtex4}

\usepackage{epsfig}
\usepackage{bm}
\usepackage{amssymb}
\usepackage{amsmath}
\usepackage{color}
\usepackage{subfigure}
\usepackage{dcolumn}

\usepackage[utf8]{inputenc}

\usepackage{amsmath}
\usepackage{amssymb}
\usepackage{amsthm}
\usepackage{amscd, amsfonts, mathrsfs}
\usepackage{cases}
\usepackage{verbatim} 
\usepackage{epsf}
\usepackage[bookmarksnumbered,pdfpagelabels=true,plainpages=false,colorlinks=true,
            linkcolor=black,citecolor=black,urlcolor=black]{hyperref}

\theoremstyle{plain}
\newtheorem{theorem}{Theorem}

\theoremstyle{definition}

\allowdisplaybreaks

\newcommand{\cA}{\mathcal A}
\newcommand{\cB}{\mathcal B}
\newcommand{\cC}{\mathcal C}

\newcommand{\cI}{\mathcal I}

\newcommand{\cM}{\mathcal M}
\newcommand{\cN}{\mathcal N}

\newcommand{\cP}{\mathcal P}
\newcommand{\cQ}{\mathcal Q}

\newcommand{\cZ}{\mathcal Z}

\newcommand{\ep}{\varepsilon}

\newcommand{\RR}{\mathbb R}

\newcommand{\cqd}{\hfill $\qed$\\ \medskip}

\newcommand{\be}{\begin{equation}}
\newcommand{\ee}{\end{equation}}
\newcommand{\bea}{\begin{eqnarray}}
\newcommand{\eea}{\end{eqnarray}}

\begin{document}


\title{Causality of the Einstein-Israel-Stewart Theory with Bulk Viscosity}
\date{\today}

\author{F\'abio S.\ Bemfica}
\affiliation{Escola de Ci\^encias e Tecnologia, Universidade Federal do Rio Grande do Norte, 
Natal, RN, Brazil (currently visiting Vanderbilt University)}
\email{fabio.bemfica@ect.ufrn.br}

\author{Marcelo M.\ Disconzi}
\affiliation{Department of Mathematics,
Vanderbilt University, Nashville, TN, USA}
\email{marcelo.disconzi@vanderbilt.edu}

\author{Jorge Noronha}
\affiliation{Instituto de F\'isica, Universidade de S\~ao Paulo, 
S\~ao Paulo, SP, Brazil}
\email{noronha@if.usp.br}


\begin{abstract}
We prove that Einstein's equations coupled to equations of Israel-Stewart-type, describing the dynamics of a relativistic fluid with bulk viscosity and nonzero baryon charge (without shear viscosity or baryon diffusion) 
dynamically coupled to gravity, are causal in the full nonlinear regime. We also show that these equations can be written as a first-order symmetric hyperbolic system, implying local existence and uniqueness of solutions to the equations of motion. 
We use an arbitrary equation of state and do not make any simplifying symmetry or near-equilibrium assumption,
requiring only physically natural conditions on the fields.
These results pave the way for the inclusion of bulk viscosity effects in simulations of gravitational-wave signals coming from 
neutron star mergers.
\end{abstract}



\maketitle


\textbf{1. Introduction.} The recent detection of a binary neutron star merger using gravitational waves \cite{TheLIGOScientific:2017qsa} and electromagnetic signals \cite{Monitor:2017mdv} marked the dawn of the multi-messenger astronomy era \cite{GBM:2017lvd}. Such events are expected to provide key information about the properties of matter at extreme densities and temperatures \cite{Abbott:2018exr}, as the density of the inner region of the object left over after the merger can be several times larger than the nuclear saturation density ($\sim 0.16$ fm$^{-3}$) while still subject
 to temperatures of the order of tens of MeV. 

Even though the properties of the equation of state of the highly dense matter formed after the merger are still uncertain \cite{Most:2018eaw}, for many years it was assumed that this system could be reasonably described as an ideal fluid (coupled to Einstein's equations) since the time scales for viscous transport to set in were previously determined \cite{Bildsten:1992my} to be outside the ten millisecond range, which is the typical time scale associated with damping due to gravitational wave emission. These estimates were recently revisited in \cite{Alford:2017rxf} using state-of-the-art merger simulations where it was concluded that, while neutrino-driven thermal transport and shear dissipation remain unlikely to affect the post-merger gravitational wave signal (unless turbulent motion occurs), damping of high-amplitude oscillations due to bulk viscosity is likely to be relevant if direct Urca processes remain suppressed. Therefore, bulk viscosity is expected to play an important role in gravitational wave emission and, as such, it should be taken into account and thoroughly investigated in merger simulations.   

However, as stressed in \cite{Alford:2017rxf}, 
the effects of bulk viscosity have not yet been included in merger simulations because this requires a formulation of relativistic fluid dynamics including bulk viscosity that is compatible with the underlying causality property of relativity theory in the strong nonlinear regime probed by the mergers (see also Refs.\ \cite{Duez_et_al_viscosity_2004,Shibata:2017xht,Radice:2018ghv} for further discussions regarding viscous effects).  

In this letter we solve this issue by proving that the equations of motion of Israel-Stewart (IS) type \cite{MIS-1,MIS-2,MIS-6}, describing a relativistic fluid with bulk viscosity and baryon charge (without shear viscosity or baryon diffusion) dynamically coupled to gravity, are causal in the full nonlinear regime when
\be
\left[\frac{\zeta}{\tau_\Pi} + n \left(\frac{\partial P}{\partial n}\right)_\varepsilon\right]\frac{1}{\varepsilon+P+\Pi} \leq 1- \left(\frac{\partial P}{\partial \varepsilon}\right)_n,
\label{maisfodao}
\ee 
where $\zeta = \zeta(\varepsilon,n)$ is the bulk viscosity, 
$\tau_\Pi=\tau_\Pi(\ep,n)$ is the bulk relaxation time, $\Pi$ is the bulk scalar, $\varepsilon$ is the energy density, $P=P(\varepsilon,n)$ is the equilibrium pressure defined by the equation of state, and $n$ is the baryon density. Requirement \eqref{maisfodao} is a simple nonlinear generalization of the known condition ensuring the linear stability of the IS equations around equilibrium. In fact, we note that near equilibrium, where  $|\Pi/(\varepsilon+P)| \ll 1$, and at zero baryon density where the speed of sound squared is $c_s^2 = dP/d\varepsilon$, \eqref{maisfodao} reduces to $
\frac{\zeta}{\tau_\Pi(\varepsilon+P)}  + \mathcal{O}\left(\frac{\Pi}{\varepsilon+P}\right)\leq  1 - c_s^2$, which is the standard condition for causality and stability in the linearized regime around equilibrium \cite{Denicol:2008ha,Pu:2009fj,Romatschke:2009im}.

We also show how to express the equations of motion of this theory that describes a bulk viscous relativistic fluid coupled to gravity as a  first-order symmetric hyperbolic (FOSH) system. This immediately implies local existence and uniqueness of its solutions, 
and sets the
equations in a form for which known numerical algorithms can be applied 
\cite{MR3235981,Reula_et_al-CausalStatistical}.
We stress that our results remain valid if one considers solely the fluid dynamic equations in a fixed (e.g., Minkowski) background.
To the best of our knowledge, this is the first time that such statements (causality, local existence, uniqueness) have been proven for IS-like theories in the nonlinear regime. 


\vskip 0.1cm
\textbf{2. The equations of motion.} IS theories of relativistic fluid dynamics \cite{MIS-1,MIS-2,MIS-6} were proposed decades ago as a way to solve the long-standing acausality and instability problem of the relativistic Navier-Stokes (NS) equations \cite{Hiscock_Lindblom_instability_1985,PichonViscous}. The basic idea is that a dissipative current such as the bulk scalar $\Pi$ does not instantaneously take its NS form\footnote{We use units $c = \hbar = k_B = 1$. The spacetime metric signature is $(-+++)$. Greek indices run from 0 to 3, Latin indices from 1 to 3.} $\Pi_{NS} = -\zeta \nabla_\mu u^\mu$ (where $u_\mu$ is the fluid 4-velocity which obeys $u_\mu u^\mu = -1$) during the fluid evolution; rather, $\Pi$ obeys a relaxation-type equation that describes how it may relax to  $\Pi_{NS}$ within the relaxation time scale $\tau_\Pi$. While such theories were originally \cite{MIS-2} derived by imposing the second law of thermodynamics for a judicious choice of the out-of-equilibrium entropy density, their modern versions in Refs.\ \cite{Baier:2007ix} and \cite{Denicol:2012cn} have focused on different aspects of relativistic hydrodynamics. Ref. \cite{Baier:2007ix} emphasized the effective field theory (EFT) character of relativistic hydrodynamics and its applicability in the strong coupling regime (see  \cite{Finazzo:2014cna} for a detailed discussion), elucidating the role played by conformal invariance and how that requires the presence of additional terms in the equations of motion that were not usually taken into account in the original IS theory. In Ref.\ \cite{Denicol:2012cn} a new moment expansion in relativistic kinetic theory \cite{kremer} was employed, together with a power counting scheme involving Knudsen and inverse Reynolds numbers, to derive the equations of motion of hydrodynamics and obtain their corresponding transport coefficients. These new values for the transport coefficients led to an overall improvement with respect to the original IS theory when comparing hydrodynamic calculations to exact solutions of the Boltzmann equation  \cite{Denicol:2012vq,Denicol:2014xca,Denicol:2014tha}.

The aforementioned versions of the IS theory differ from their original counterpart
while retaining its basic physical insights.  As described in the previous paragraph, 
there are different theories representing an improved, albeit different, formulation
of the IS equations. These theories will be referred to here as generalized IS theories.

With applications to neutron star mergers in mind, here we only consider the dissipative effects coming from bulk viscosity. The fluid energy-momentum tensor is given by $T^{\mu\nu} = \varepsilon u^\mu u^\nu + (P+\Pi) \Delta^{\mu\nu}$, where $\Delta_{\mu\nu} = g_{\mu\nu} + u_\mu u_\nu$ is the projector orthogonal to $u_\mu$ and $g_{\mu\nu}$ is the spacetime metric. Using $u^\mu u_\mu = -1$, energy-momentum conservation, $\nabla_\mu T^{\mu\nu}=0$, is equivalent to the equations
\bea
 u^\alpha \nabla_\alpha \varepsilon + \left(\varepsilon+P+\Pi\right)\nabla_\alpha u^\alpha & = 0 \label{conservaEM1} ,\\
 \left(\varepsilon+P+\Pi \right)u^\beta \nabla_\beta u_\alpha + \alpha_1\Delta^{\beta}_\alpha \nabla_\beta  \varepsilon 
 \nonumber
 \\
 + \alpha_2\Delta^{\beta}_\alpha \nabla_\beta n+  \Delta^{\beta}_\alpha \nabla_\beta \Pi & =0
\label{conservaEM2},
\eea
where 
$\alpha_1 = \left( \partial P/\partial \varepsilon \right)_n$ and
$\alpha_2 = \left( \partial P/\partial n \right)_\varepsilon$. 
Eqs. \eqref{conservaEM1}--\eqref{conservaEM2} are supplemented by the following bulk scalar relaxation equation
\be
\tau_\Pi u^\alpha \nabla_\alpha \Pi + \Pi + \lambda \Pi^2+ \zeta \nabla_\alpha u^\alpha=0,
\label{ISbulk}
\ee
where $\lambda = \lambda(\varepsilon,n)$ is a transport coefficient. 
Eq. \eqref{ISbulk} corresponds to Eq. (63) in \cite{Denicol:2012cn} without
shear viscosity or baryon diffusion and with $\delta_{\Pi \Pi} =0$. It is common
practice to omit the term proportional to $(\nabla_\alpha u^\alpha)^2$ in \cite{Denicol:2012cn} and, therefore,
we have also done so here \cite{Ryu:2015vwa}. We choose to work with this specific generalized IS equation because it contains all the relevant
physics while making the interpretation of our results clear. In fact, although it is possible to include other terms in the bulk channel \cite{Denicol:2012cn,Romatschke:2017ejr,Philipsen:2013nea}, the above equation contains the essential terms for our discussion: $\tau_\Pi$ is associated with causality, $\lambda$ parametrizes the presence of nonlinear terms that do not contain derivatives of the fields, and the term $ \zeta \nabla_\alpha u^\alpha$ ensures that the NS limit can be recovered.
In Sec. 2.3
we explain how our methods apply with no change to other generalized IS theories.

We also include a non-zero baryon current $J_\mu = n\, u_\mu$, whose conservation gives  (our results can be easily adapted for 
$n=0$)
\be
u^\mu \nabla_\mu n + n \nabla_\mu u^\mu = 0.
\label{baryoneq}
\ee
The fluid is dynamically coupled to gravity via Einstein's equations
\be
R_{\mu\nu} - \frac{1}{2}R g_{\mu\nu} +\Lambda g_{\mu\nu} = 8 \pi G \,T_{\mu\nu},
\label{einstein}
\ee
where $R_{\mu\nu}$ is the Ricci tensor, $R = g_{\mu\nu}R^{\mu\nu}$, $\Lambda$ is the cosmological constant, and $G$ is Newton's  gravitational constant. 
Eqs. \eqref{conservaEM1}--\eqref{einstein},
with $u^\mu u_\mu = -1$, define the generalized Einstein-Israel-Stewart
(EIS) theory considered in this work.


\vskip 0.1cm
\emph{2.1. Causality.}
Causality is the idea that no information propagates faster than the speed of light
and that no closed timelike curves exist in spacetime, i.e., the future cannot influence 
the past (see the Appendix for the precise definition of causality).
This concept lies at the core of relativity theory and, therefore,
the dynamics of the fluid sector must be compatible with it.
Despite its importance, causality has not been established in the nonlinear regime 
for the IS theory or its variations (see \cite{Bemfica:2017wps,DisconziKephartScherrerNew,RezzollaZanottiBookRelHydro}
for a discussion).
While causality is known to hold in ideal relativistic hydrodynamics at the nonlinear level 
\cite{Choquet-BruhatFluidsExistence, ChoquetBruhatGRBook, DisconziRemarksEinsteinEuler,Lichnerowicz_MHD_book}, only statements valid in the linearized regime around equilibrium are known for the IS theory \cite{Hiscock_Lindblom_stability_1983,Hiscock_Lindblom_instability_1985,Olson:1989ey,Denicol:2008ha,Pu:2009fj}.
Attempts to go beyond the linear regime, although restricted to $1+1$ dimensions
or assuming very strong symmetry conditions, and in flat spacetime, appeared, respectively, in \cite{Denicol:2008ha} and
\cite{Floerchinger:2017cii} (compare \eqref{maisfodao} with the causality
condition found in \cite{Denicol:2008ha}).
Thus, a general proof of causality of IS-like systems (coupled to gravity) is so far lacking.

We show that if $\frac{\zeta}{\tau_\Pi(\varepsilon+P+\Pi)}+\alpha_1 
+ \frac{\alpha_2 \,n}{\varepsilon + P+\Pi} \geq 0$ and condition \eqref{maisfodao} is satisfied,
then the generalized EIS system is causal. Causality of Eqs. \eqref{conservaEM1}--\eqref{baryoneq} in a fixed background also holds
under the same assumptions.

We refer to Theorem 1 
in the Appendix for 
a formal statement of the causality of the EIS system as well as its proof.
In a nutshell, in the Appendix we establish that the values 
of the fields $\ep,u_\alpha, \Pi$, and $g_{\alpha \beta}$ at a point $p$ are influenced
only by the dynamics of such fields in the causal past of $p$. In Minkowski space, 
the causal past of $p$ is simply the bottom half of the lighcone with vertex at $p$.
This generalizes to curved spacetimes, where the half-bottom of the light cone is replaced by a curved cone-like region with vertex at $p$.

Condition $\frac{\zeta}{\tau_\Pi(\varepsilon+P+\Pi)}+\alpha_1 
+ \frac{\alpha_2 \,n}{\varepsilon + P+\Pi} \geq 0$ is physically very natural as $\alpha_1 
+ \frac{\alpha_2 \,n}{\varepsilon + P}$ is the speed of sound squared in equilibrium and $\zeta$ is
non-negative. 

\vskip 0.1cm
\emph{2.2. The EIS equations as a FOSH system, existence, and uniqueness.} 
A system of first order partial differential equations for an unknown vector 
$\Phi$ is said to be a FOSH system if it can be written as
$\cA^\mu(\Phi) \partial_\mu \Phi + \cB(\Phi) = F$, where $\cA^\mu$ are matrices and
$\cB$ is a vector, all possibly depending on $\Phi$ but not on its derivatives, 
$\cA^\mu$ are symmetric, $\cA^0$ is positive definite, and $F$
is a given source term.

Several important properties are readily available for FOSH
systems \cite{Courant_and_Hilbert_book_2,Geroch:1996kg,Frittelli:1996wr}. 
One such property is that for these systems the initial-value problem admits
existence and uniqueness of solutions. Besides assuring a  
firm theoretical basis for the system, knowing that the equations admit existence
and uniqueness of solutions is helpful to ensure the reliability of
many numerical schemes; see \cite{GuermondetalNumerical}
for examples of the potential pitfalls of simulating equations for which no existence and uniqueness
results are available, and \cite{Bemfica:2017wps} for a complementary discussion. While
for most of standard physical theories existence and uniqueness of solutions had long been
settled \cite{ChoquetBruhatGRBook}, for theories of relativistic fluids with viscosity very 
few results are available \cite{Bemfica:2017wps,DisconziViscousFluidsNonlinearity,DisconziFollowupBemficaNoronha},
and none so far for the generalized IS theory in the nonlinear regime. 

Eqs \eqref{conservaEM1}--\eqref{baryoneq} can be written as a FOSH system with
$\Phi = (\varepsilon,u^i, n, \Pi)$, a suitable $\cB$, 
$F=0$, and $\cA^0$ and $\cA^k$ given, respectively, by
\bea
\begin{pmatrix}
\frac{u^0}{(\varepsilon+P+\Pi)} & -\frac{u_i}{u_0} & 0 & 0 \\
  -\frac{u_j}{u_0} & A_{ij}u^0\frac{\left(\varepsilon+P+\Pi\right)}{\left(\frac{\partial P}{\partial \varepsilon}\right)_n} &-\frac{u_j}{u_0}\frac{\left(\frac{\partial P}{\partial n}\right)_\varepsilon}{\left(\frac{\partial P}{\partial \varepsilon}\right)_n} & -\frac{u_j}{u_0}\frac{1}{\left(\frac{\partial P}{\partial \varepsilon}\right)_n}  \\
  0 & -\frac{u_i}{u_0} \frac{\left(\frac{\partial P}{\partial n}\right)_\varepsilon}{\left(\frac{\partial P}{\partial \varepsilon}\right)_n}& \frac{u^0}{n}\frac{\left(\frac{\partial P}{\partial n}\right)_\varepsilon}{\left(\frac{\partial P}{\partial \varepsilon}\right)_n}& 0  \\
    0 & -\frac{u_i}{u_0}\frac{1}{\left(\frac{\partial P}{\partial \varepsilon}\right)_n}& 0 & \frac{\tau_\Pi}{\zeta \left(\frac{\partial P}{\partial \varepsilon}\right)_n} u^0 & \\
 \end{pmatrix}
 \nonumber
\eea
and\linebreak
\bea
 \begin{pmatrix}
 \frac{u^k}{\left(\varepsilon+P+\Pi\right)} & \delta^k_i & 0 & 0 \\
  \delta^k_j & A_{ij}u^k\frac{\left(\varepsilon+P+\Pi\right)}{\left(\frac{\partial P}{\partial \varepsilon}\right)_n} & \delta_j^k \frac{\left(\frac{\partial P}{\partial n}\right)_\varepsilon}{\left(\frac{\partial P}{\partial \varepsilon}\right)_n} & \frac{\delta_j^k}{\left(\frac{\partial P}{\partial \varepsilon}\right)_n} \\
    0 & \delta_i^k \frac{\left(\frac{\partial P}{\partial n}\right)_\varepsilon}{\left(\frac{\partial P}{\partial \varepsilon}\right)_n} & \frac{u^k}{n}\frac{\left(\frac{\partial P}{\partial n}\right)_\varepsilon}{\left(\frac{\partial P}{\partial \varepsilon}\right)_n} & 0\\
  0 & \frac{\delta_i^k}{\left(\frac{\partial P}{\partial \varepsilon}\right)_n} & 0 &  \frac{\tau_\Pi}{\zeta\left(\frac{\partial P}{\partial \varepsilon}\right)_n} u^k  
 \end{pmatrix},
\nonumber
\eea
where $A_{ij} = g_{ij} - g_{i0}\frac{u_j}{u_0} - \frac{u_i}{u_0}g_{0j}+g_{00}\frac{u_i u_j}{u_0^2}$.

Using the above formulation of the generalized IS equations as a FOSH, we establish existence and uniqueness
of solutions to the initial-value formulation for the generalized EIS system under suitable assumptions.
More precisely, given an equation of state $P = P(\varepsilon, n)$,
a bulk viscosity $\zeta = \zeta(\varepsilon,n)$, and a relaxation time $\tau_\Pi=\tau_\Pi(\varepsilon,n)$, consider initial conditions
$\mathring{\ep} \equiv \ep(0), \mathring{u}^i \equiv u^i(0), 
\mathring{\Pi} \equiv \Pi(0), \mathring{n} \equiv n(0), 
 g_{ij}(0), \partial_0 g_{ij}(0)$ satisfying \eqref{maisfodao} 
 along an initial surface
$\Sigma \equiv \{ x^0 = 0 \}$.  Assume that $\mathring{\varepsilon}+P(\mathring{\varepsilon}, \mathring{n})+\mathring{\Pi}$, $\tau_\Pi (\mathring{\varepsilon}, \mathring{n})$, $\zeta(\mathring{\varepsilon}, \mathring{n})>0$, $\frac{\partial P}{\partial \varepsilon}(\mathring{\varepsilon},\mathring{n}) + \frac{\partial P}{\partial n}(\mathring{\varepsilon},\mathring{n}) \mathring{n}/(\mathring{\varepsilon}+P(\mathring{\varepsilon}, \mathring{n})+\mathring{\Pi})\ge 0$, 
and that   
$\mathring{n}$, $\frac{\partial P}{\partial \varepsilon}(\mathring{\varepsilon},\mathring{n})$, and $\frac{\partial P}{\partial n}(\mathring{\varepsilon},\mathring{n})$ are nonzero. Then there exists
a spacetime $M$ containing $\Sigma$ such that a unique solution to the generalized EIS equations exists
on $M$. A similar statement, i.e., existence and uniqueness of solutions 
for initial conditions as above, holds for the generalized IS equations in a fixed background.

We refer to Theorem 2 in the Appendix for a formal statement of existence
and uniqueness of solutions, its proof, and for the generalized IS as a FOSH.
Here, we make some relevant remarks.

Assumptions
$\mathring{\varepsilon}+P(\mathring{\varepsilon}, \mathring{n})+\mathring{\Pi}$, $\tau_\Pi (\mathring{\varepsilon}, \mathring{n})$, $\zeta(\mathring{\varepsilon}, \mathring{n})>0$, and $\frac{\partial P}{\partial \varepsilon}(\mathring{\varepsilon},\mathring{n}) + \frac{\partial P}{\partial n}(\mathring{\varepsilon},\mathring{n}) \mathring{n}/(\mathring{\varepsilon}+P(\mathring{\varepsilon}, \mathring{n})+\mathring{\Pi})\ge 0$ are physically natural (our results are easily adapted to the case when $n$ is absent), while
$\frac{\partial P}{\partial \varepsilon}(\mathring{\varepsilon},\mathring{n})$, $\frac{\partial P}{\partial n}(\mathring{\varepsilon},\mathring{n}) \neq 0$ are very mild requirements. 
Note that these can be viewed as conditions on $\mathring{\varepsilon}$ and $\mathring{n}$,
with the form of the equation of state remaining rather general.
It is well-known that because of the covariance of Einstein's equations
only $g_{ij}$ and $\partial_0 g_{ij}$, rather than $g_{\alpha\beta}$ and $\partial_0 g_{\alpha\beta}$, 
are given as initial data \cite{HawkingEllisBook,WaldBookGR1984}. Similarly, only a vector field on $\Sigma$, i.e., $\mathring{u}^i$ rather
than $\mathring{u}^\alpha$, is given along $\Sigma$, with the full four-velocity on $\Sigma$
determined from $u^\mu u_\mu = -1$. 

The proof of Theorem 2 follows from known arguments
once the generalized IS equations are formulated as a FOSH system. The main difficulty to achieve the latter is
that there is no method to determine whether a given set of equations can in principle be written
as a FOSH system. Here, we relied on the following two ingredients to achieve this.
First, computing the characteristics of the system (which is used for establishing the causality of the
equations), we find them to be a combination of the flow lines 
of $u^\alpha$ and of null-cones with respect to an acoustical metric. The former is associated 
with transport equations (which are the prototypical example of a FOSH system); the latter
is associated with wave equations (which can be rewritten as FOSH). This suggests
that a FOSH formulation might be possible.
Second, Eqs.\ \eqref{conservaEM1}, \eqref{conservaEM2}, and \eqref{baryoneq}
resemble the equations
of an ideal fluid, for which a FOSH formulation is known \cite{AnileBook,AnilePennisiMHD}.
This suggests trying to adapt the procedure that works for an ideal fluid. In this regard,
it is crucial that the viscous contributions due to bulk viscosity are given by a scalar:
if vector (baryon diffusion) or tensor (shear viscosity) viscous contributions are included, 
the characteristics become very complicated and Eqs.\ \eqref{conservaEM1}, \eqref{conservaEM2}, and \eqref{baryoneq} no longer resemble those of an ideal fluid. It is not clear whether
causality (in the nonlinear regime) or a formulation as a FOSH system can be obtained in these cases 
following the approach pursued here.

We assumed that all transport coefficients depend only on $\varepsilon$ and $n$
(however, see Sec. 2.3). 
Transport coefficients also generally depend on some characteristic microscopic variables
\cite{Denicol:2011fa}. In many applications, however, these microscopic
variables are either neglected, treated as parameters that can be independently
estimated and thus considered given (such as the viscosities in Navier-Stokes theory), or directly eliminated. Otherwise,
 one would not be dealing with a purely hydrodynamic 
theory \cite{Romatschke:2017ejr}. Therefore, our assumptions fall well within the scope of applicability
of hydrodynamic models.
See also Sec. 2.3 for a more general setting where the dependence on the microscopic
dynamics can be decoupled upon simple assumptions.

\vskip 0.1cm
\emph{2.3. Other theories.} Here we point out that our methods give, with no change,
causality, existence, and uniqueness of solutions to Einstein's equations coupled
to other generalized IS theories (without shear viscosity or baryon diffusion). A quick inspection 
in our proofs provided in the Appendix reveals that they remain true, with no change,
if $\tau_\Pi$ and $\zeta$ are allowed to depend on $\Pi$. In particular, consider Eq. (63) in \cite{Denicol:2012cn} 
with $\delta_{\Pi \Pi}$ not necessarily zero. Defining $\bar{\zeta} = \zeta + 
\delta_{\Pi \Pi} \Pi$, we obtain Eq. \eqref{ISbulk} with $\zeta$ replaced by $\bar{\zeta}$.
Hence, all our results go through with $\bar{\zeta}$ in place of $\zeta$. Our assumptions in this
case, however, seem less natural since, for example, there is no reason to expect 
$\bar{\zeta}$ to be positive.

Finally, the methods we employ depend only on the principal part of the system\footnote{I.e., 
the subsystem comprised only of the terms involving derivatives in the equations.}. Thus
all our conclusions remain valid if Eq. \eqref{ISbulk} is replaced by
\be
\tau_\Pi u^\alpha \nabla_\alpha \Pi + \zeta \nabla_\alpha u^\alpha +f(\ep, n, \Pi, \tau_\Pi, \zeta)
=0,
\nonumber
\ee
where $f$ is an arbitrary function of $\ep$, $n$, $\Pi$, $\tau_\Pi$, and $\zeta$ 
(but not depending on their derivatives) and $\tau_\Pi$ and $\zeta$ are allowed to depend on
$\Pi$ (so that, in particular, $\zeta$ can be replaced by $\bar{\zeta}$ here as well).

Consider now the case when $\zeta$ and $\tau_\Pi$ depend on a microscopic
variable $\ell$, and suppose that $\ell$ cannot be neglected or treated as a parameter
but needs to be found by solving its own evolution equation which couples to the
hydrodynamic variables. If one assumes, as it is often
done \cite{Ryu:2015vwa}, that the ratio $\zeta/\tau_\Pi$
depends only $\ep$, $n$, and $\Pi$, then all our conclusions remain valid as long
as $f$ is such that
$\frac{1}{\tau_\Pi} f(\ep, n, \Pi, \tau_\Pi, \zeta)$ does not involve dynamical microscopic variables.
Thus, our results are applicable to many cases of interest in heavy ion collisions (e.g., \cite{Denicol:2018wdp}) and neutron star mergers. In this regard, note that our condition
\eqref{maisfodao} depends only on the ratio $\zeta/\tau_\Pi$.

\vskip 0.1cm
\textbf{3. Conclusions.} In this paper we proved that the generalized EIS equations for a relativistic fluid with bulk viscosity and nonzero baryon charge (without shear viscosity or baryon diffusion) 
are causal in the full nonlinear regime. We also proved that these equations admit a FOSH formulation, existence, and uniqueness of solutions. Our results
hold under hypothesis typically satisfied by viscous relativistic fluids.
In particular, we do not require any symmetry or near-equilibrium assumption.
The most immediate
application of these results is in the study of neutron star mergers.
In this regard, it is crucial that our results hold under general assumptions on the  equation of state since the latter is not yet known from first principles.

Given that theories of IS type are the most widely used formulation of relativistic viscous fluid dynamics  \cite{Jeon:2015dfa}, it is crucial to put them on a firm theoretical (and mathematical) foundation. Combining our results with known properties of the 
generalized IS equations \cite{MIS-6,Pu:2009fj,Romatschke:2009im,Denicol:2012cn}, the case considered here seems to be the first example in the literature of a theory of relativistic viscous fluids such that (i) causality, local existence, and uniqueness
of solutions hold in the nonlinear regime with or without coupling to Einstein's equations; 
(ii) linear stability around equilibrium holds; (iii) the equations of motion are derivable from microscopic approaches (such as kinetic theory); 
and (iv) solutions are guaranteed to exist in function spaces (the Sobolev spaces,
see the Appendix) well-suited for the implementation
of numerical codes. (The theory introduced in \cite{Bemfica:2017wps} also satisfies 
(i)-(iii), but it is applicable only to conformal fluids and its solutions exist on 
function spaces more restrictive than Sobolev spaces.) This is the first time that all such properties are shown to hold since Eckart's seminal work in 1940 \cite{EckartViscous}.

As Theorems 1 and 2 settle the basic question of causality and existence of solutions to the EIS system,
it is now possible to generalize to relativistic viscous fluids key results
known to hold for ideal fluids, such as the formation of shocks \cite{ChristodoulouShocks},
global-in-time results \cite{SpeckNonlinearStability}, or the problem of accurately 
describing the fluid-vacuum interface on stars \cite{RendallFluidBodies}.

These results have two further applications: the hydrodynamic evolution of the quark-gluon plasma formed in heavy ion collisions \cite{Heinz:2013th} and cosmology. Current state-of-the-art modeling of quark-gluon plasma dynamics involves solving IS-like equations taking into account shear and bulk viscosities and, more recently, baryon diffusion \cite{Denicol:2018wdp}. Differently than the case of neutron star mergers \cite{Alford:2017rxf}, in heavy ion collisions shear and baryon diffusion effects are not negligible though bulk viscosity also plays an important role \cite{Noronha-Hostler:2013gga,Noronha-Hostler:2014dqa,Ryu:2015vwa}. In this regard, we note 
Theorem 2 
guarantees existence of solutions even in the presence of cavitation where $P + \Pi \sim 0$ \cite{Torrieri:2007fb,Rajagopal:2009yw} and $\varepsilon >0$, a phenomenon whose numerical description can be quite challenging \cite{Denicol:2015bpa}. Another application of the EIS system studied here is in cosmology, where the inclusion of bulk viscosity has been widely studied \cite{MaartensDissipative, LiBarrow, Disconzi_Kephart_Scherrer_2015, Zimdahl, Piattella_et_al}.
Indeed, our results do not make any symmetry assumptions, thus allowing the study of cosmological models with viscosity outside the symmetry class of homogeneous and isotropic solutions (e.g. \cite{MontaniLichnerowiczViscosityBianchi,East:2015ggf}). 

The equations here investigated suffice for many important applications (including
neutron star mergers).
Recent developments in EFTs, however, indicate that a more
complete description of fluid phenomena should include further terms in the equations.
While such contributions are negligible in many cases of interest, 
they consist of a genuine prediction of EFT, and 
their omission 
can lead to inconsistencies in 2${}^\text{nd}$ order Kubo formulas
\cite{Romatschke:2017ejr,Philipsen:2013nea}. 
Generalizing our results to theories that include these effects is, therefore, 
important for a more complete description of relativistic viscous fluids. While
our techniques do not apply directly to such cases, we expect them to provide 
a starting point for studying these scenarios.

Our results open the door for new in-depth studies of fluid dynamics under extreme conditions such as in relativistic turbulent phenomena \cite{Eyink:2017zfz} and neutron star mergers. In particular, the latter is expected to take center stage in the coming years and our causality and existence results, which remain valid in the full nonlinear regime, provide the necessary cornerstone for quantitative studies of non-equilibrium viscous fluid phenomena in strong gravitational fields.


\vskip 0.1cm
\textbf{Acknowledgements.} We would like to thank the anonymous referees for 
a careful reading of the manuscript and several comments that improved the paper.
FSB is partially supported by 
a Discovery Grant administered by Vanderbilt University. MMD is partially supported by a Sloan Research Fellowship provided by the Alfred P. Sloan foundation, NSF grant DMS-1812826, and a Discovery Grant administered by Vanderbilt University. JN is partially supported by 
CNPq grant 306795/2017-5 and FAPESP grant 2017/05685-2. 

\bibliography{References.bib}

\def\cprime{$'$}
\begin{thebibliography}{76}
\expandafter\ifx\csname natexlab\endcsname\relax\def\natexlab#1{#1}\fi
\expandafter\ifx\csname bibnamefont\endcsname\relax
  \def\bibnamefont#1{#1}\fi
\expandafter\ifx\csname bibfnamefont\endcsname\relax
  \def\bibfnamefont#1{#1}\fi
\expandafter\ifx\csname citenamefont\endcsname\relax
  \def\citenamefont#1{#1}\fi
\expandafter\ifx\csname url\endcsname\relax
  \def\url#1{\texttt{#1}}\fi
\expandafter\ifx\csname urlprefix\endcsname\relax\def\urlprefix{URL }\fi
\providecommand{\bibinfo}[2]{#2}
\providecommand{\eprint}[2][]{\url{#2}}

\bibitem[{\citenamefont{Abbott
  et~al.}(2017{\natexlab{a}})}]{TheLIGOScientific:2017qsa}
\bibinfo{author}{\bibfnamefont{B.}~\bibnamefont{Abbott}} \bibnamefont{et~al.}
  (\bibinfo{collaboration}{LIGO Scientific, Virgo}), \bibinfo{journal}{Phys.
  Rev. Lett.} \textbf{\bibinfo{volume}{119}}, \bibinfo{pages}{161101}
  (\bibinfo{year}{2017}{\natexlab{a}}), \eprint{1710.05832}.

\bibitem[{\citenamefont{Abbott et~al.}(2017{\natexlab{b}})}]{Monitor:2017mdv}
\bibinfo{author}{\bibfnamefont{B.~P.} \bibnamefont{Abbott}}
  \bibnamefont{et~al.} (\bibinfo{collaboration}{LIGO Scientific, Virgo,
  Fermi-GBM, INTEGRAL}), \bibinfo{journal}{Astrophys. J.}
  \textbf{\bibinfo{volume}{848}}, \bibinfo{pages}{L13}
  (\bibinfo{year}{2017}{\natexlab{b}}), \eprint{1710.05834}.

\bibitem[{\citenamefont{Abbott et~al.}(2017{\natexlab{c}})}]{GBM:2017lvd}
\bibinfo{author}{\bibfnamefont{B.~P.} \bibnamefont{Abbott}}
  \bibnamefont{et~al.} (\bibinfo{collaboration}{LIGO Scientific, Virgo, Fermi
  GBM, INTEGRAL, and others}), \bibinfo{journal}{Astrophys. J.}
  \textbf{\bibinfo{volume}{848}}, \bibinfo{pages}{L12}
  (\bibinfo{year}{2017}{\natexlab{c}}), \eprint{1710.05833}.

\bibitem[{\citenamefont{Abbott et~al.}(2018)}]{Abbott:2018exr}
\bibinfo{author}{\bibfnamefont{B.~P.} \bibnamefont{Abbott}}
  \bibnamefont{et~al.} (\bibinfo{collaboration}{LIGO Scientific, Virgo}),
  \bibinfo{journal}{Phys. Rev. Lett.} \textbf{\bibinfo{volume}{121}},
  \bibinfo{pages}{161101} (\bibinfo{year}{2018}), \eprint{1805.11581}.

\bibitem[{\citenamefont{Most et~al.}(2019)\citenamefont{Most, Papenfort,
  Dexheimer, Hanauske, Schramm, St{\"o}cker., and Rezzolla}}]{Most:2018eaw}
\bibinfo{author}{\bibfnamefont{E.~R.} \bibnamefont{Most}},
  \bibinfo{author}{\bibfnamefont{L.~J.} \bibnamefont{Papenfort}},
  \bibinfo{author}{\bibfnamefont{V.}~\bibnamefont{Dexheimer}},
  \bibinfo{author}{\bibfnamefont{M.}~\bibnamefont{Hanauske}},
  \bibinfo{author}{\bibfnamefont{S.}~\bibnamefont{Schramm}},
  \bibinfo{author}{\bibfnamefont{H.}~\bibnamefont{St{\"o}cker.}},
  \bibnamefont{and} \bibinfo{author}{\bibfnamefont{L.}~\bibnamefont{Rezzolla}},
  \bibinfo{journal}{Phys. Rev. Lett.} \textbf{\bibinfo{volume}{122}},
  \bibinfo{pages}{061101} (\bibinfo{year}{2019}).

\bibitem[{\citenamefont{Bildsten and Cutler}(1992)}]{Bildsten:1992my}
\bibinfo{author}{\bibfnamefont{L.}~\bibnamefont{Bildsten}} \bibnamefont{and}
  \bibinfo{author}{\bibfnamefont{C.}~\bibnamefont{Cutler}},
  \bibinfo{journal}{Astrophys. J.} \textbf{\bibinfo{volume}{400}},
  \bibinfo{pages}{175} (\bibinfo{year}{1992}).

\bibitem[{\citenamefont{Alford et~al.}(2018)\citenamefont{Alford, Bovard,
  Hanauske, Rezzolla, and Schwenzer}}]{Alford:2017rxf}
\bibinfo{author}{\bibfnamefont{M.~G.} \bibnamefont{Alford}},
  \bibinfo{author}{\bibfnamefont{L.}~\bibnamefont{Bovard}},
  \bibinfo{author}{\bibfnamefont{M.}~\bibnamefont{Hanauske}},
  \bibinfo{author}{\bibfnamefont{L.}~\bibnamefont{Rezzolla}}, \bibnamefont{and}
  \bibinfo{author}{\bibfnamefont{K.}~\bibnamefont{Schwenzer}},
  \bibinfo{journal}{Phys. Rev. Lett.} \textbf{\bibinfo{volume}{120}},
  \bibinfo{pages}{041101} (\bibinfo{year}{2018}), \eprint{1707.09475}.

\bibitem[{\citenamefont{Duez et~al.}(2004)\citenamefont{Duez, Liu, and
  Shapiro}}]{Duez_et_al_viscosity_2004}
\bibinfo{author}{\bibfnamefont{M.~D.} \bibnamefont{Duez}},
  \bibinfo{author}{\bibfnamefont{Y.~T.} \bibnamefont{Liu}}, \bibnamefont{and}
  \bibinfo{author}{\bibfnamefont{S.~L.} \bibnamefont{Shapiro}},
  \bibinfo{journal}{Phys. Rev. D} \textbf{\bibinfo{volume}{69}},
  \bibinfo{pages}{104030} (\bibinfo{year}{2004}).

\bibitem[{\citenamefont{Shibata and Kiuchi}(2017)}]{Shibata:2017xht}
\bibinfo{author}{\bibfnamefont{M.}~\bibnamefont{Shibata}} \bibnamefont{and}
  \bibinfo{author}{\bibfnamefont{K.}~\bibnamefont{Kiuchi}},
  \bibinfo{journal}{Phys. Rev.} \textbf{\bibinfo{volume}{D95}},
  \bibinfo{pages}{123003} (\bibinfo{year}{2017}), \eprint{1705.06142}.

\bibitem[{\citenamefont{Radice et~al.}(2018)\citenamefont{Radice, Perego,
  Hotokezaka, Bernuzzi, Fromm, and Roberts}}]{Radice:2018ghv}
\bibinfo{author}{\bibfnamefont{D.}~\bibnamefont{Radice}},
  \bibinfo{author}{\bibfnamefont{A.}~\bibnamefont{Perego}},
  \bibinfo{author}{\bibfnamefont{K.}~\bibnamefont{Hotokezaka}},
  \bibinfo{author}{\bibfnamefont{S.}~\bibnamefont{Bernuzzi}},
  \bibinfo{author}{\bibfnamefont{S.~A.} \bibnamefont{Fromm}}, \bibnamefont{and}
  \bibinfo{author}{\bibfnamefont{L.~F.} \bibnamefont{Roberts}},
  \bibinfo{journal}{Astrophys. J. Lett.} \textbf{\bibinfo{volume}{869}},
  \bibinfo{pages}{L35} (\bibinfo{year}{2018}), \eprint{1809.11163}.

\bibitem[{\citenamefont{Mueller}(1967)}]{MIS-1}
\bibinfo{author}{\bibfnamefont{I.}~\bibnamefont{Mueller}},
  \bibinfo{journal}{Zeit. fur Phys} \textbf{\bibinfo{volume}{198}},
  \bibinfo{pages}{329} (\bibinfo{year}{1967}).

\bibitem[{\citenamefont{Israel}(1976)}]{MIS-2}
\bibinfo{author}{\bibfnamefont{W.}~\bibnamefont{Israel}},
  \bibinfo{journal}{Ann. Phys.} \textbf{\bibinfo{volume}{100}},
  \bibinfo{pages}{310} (\bibinfo{year}{1976}).

\bibitem[{\citenamefont{Israel and Stewart}(1979)}]{MIS-6}
\bibinfo{author}{\bibfnamefont{W.}~\bibnamefont{Israel}} \bibnamefont{and}
  \bibinfo{author}{\bibfnamefont{J.~M.} \bibnamefont{Stewart}},
  \bibinfo{journal}{Ann. Phys.} \textbf{\bibinfo{volume}{118}},
  \bibinfo{pages}{341} (\bibinfo{year}{1979}).

\bibitem[{\citenamefont{Denicol et~al.}(2008)\citenamefont{Denicol, Kodama,
  Koide, and Mota}}]{Denicol:2008ha}
\bibinfo{author}{\bibfnamefont{G.~S.} \bibnamefont{Denicol}},
  \bibinfo{author}{\bibfnamefont{T.}~\bibnamefont{Kodama}},
  \bibinfo{author}{\bibfnamefont{T.}~\bibnamefont{Koide}}, \bibnamefont{and}
  \bibinfo{author}{\bibfnamefont{P.}~\bibnamefont{Mota}}, \bibinfo{journal}{J.
  Phys.} \textbf{\bibinfo{volume}{G35}}, \bibinfo{pages}{115102}
  (\bibinfo{year}{2008}), \eprint{0807.3120}.

\bibitem[{\citenamefont{Pu et~al.}(2010)\citenamefont{Pu, Koide, and
  Rischke}}]{Pu:2009fj}
\bibinfo{author}{\bibfnamefont{S.}~\bibnamefont{Pu}},
  \bibinfo{author}{\bibfnamefont{T.}~\bibnamefont{Koide}}, \bibnamefont{and}
  \bibinfo{author}{\bibfnamefont{D.~H.} \bibnamefont{Rischke}},
  \bibinfo{journal}{Phys. Rev.} \textbf{\bibinfo{volume}{D81}},
  \bibinfo{pages}{114039} (\bibinfo{year}{2010}), \eprint{0907.3906}.

\bibitem[{\citenamefont{Romatschke}(2010)}]{Romatschke:2009im}
\bibinfo{author}{\bibfnamefont{P.}~\bibnamefont{Romatschke}},
  \bibinfo{journal}{Int. J. Mod. Phys.} \textbf{\bibinfo{volume}{E19}},
  \bibinfo{pages}{1} (\bibinfo{year}{2010}), \eprint{0902.3663}.

\bibitem[{\citenamefont{Gustafsson et~al.}(2013)\citenamefont{Gustafsson, ,
  Kreiss, and Oliger}}]{MR3235981}
\bibinfo{author}{\bibfnamefont{B.}~\bibnamefont{Gustafsson}}, ,
  \bibinfo{author}{\bibfnamefont{H.-O.} \bibnamefont{Kreiss}},
  \bibnamefont{and} \bibinfo{author}{\bibfnamefont{J.}~\bibnamefont{Oliger}},
  \emph{\bibinfo{title}{Time-dependent problems and difference methods}}, Pure
  and Applied Mathematics (Hoboken) (\bibinfo{publisher}{John Wiley \& Sons,
  Inc., Hoboken, NJ}, \bibinfo{year}{2013}), \bibinfo{edition}{2nd} ed., ISBN
  \bibinfo{isbn}{978-0-470-90056-7},
  \urlprefix\url{https://doi.org/10.1002/9781118548448}.

\bibitem[{\citenamefont{Reula and Nagy}(1997)}]{Reula_et_al-CausalStatistical}
\bibinfo{author}{\bibfnamefont{O.~A.} \bibnamefont{Reula}} \bibnamefont{and}
  \bibinfo{author}{\bibfnamefont{G.~B.} \bibnamefont{Nagy}},
  \bibinfo{journal}{J. Phys. A} \textbf{\bibinfo{volume}{30}},
  \bibinfo{pages}{1695} (\bibinfo{year}{1997}), ISSN \bibinfo{issn}{0305-4470},
  \urlprefix\url{http://dx.doi.org/10.1088/0305-4470/30/5/030}.

\bibitem[{\citenamefont{Hiscock and
  Lindblom}(1985)}]{Hiscock_Lindblom_instability_1985}
\bibinfo{author}{\bibfnamefont{W.~A.} \bibnamefont{Hiscock}} \bibnamefont{and}
  \bibinfo{author}{\bibfnamefont{L.}~\bibnamefont{Lindblom}},
  \bibinfo{journal}{Phys. Rev. D} \textbf{\bibinfo{volume}{31}},
  \bibinfo{pages}{725} (\bibinfo{year}{1985}).

\bibitem[{\citenamefont{Pichon}(1965)}]{PichonViscous}
\bibinfo{author}{\bibfnamefont{G.}~\bibnamefont{Pichon}},
  \bibinfo{journal}{Annales de l'I.H.P. Physique th{\'e}orique}
  \textbf{\bibinfo{volume}{2}}, \bibinfo{pages}{21} (\bibinfo{year}{1965}),
  \urlprefix\url{http://www.numdam.org/item/AIHPA_1965__2_1_21_0}.

\bibitem[{\citenamefont{Baier et~al.}(2008)\citenamefont{Baier, Romatschke,
  Son, Starinets, and Stephanov}}]{Baier:2007ix}
\bibinfo{author}{\bibfnamefont{R.}~\bibnamefont{Baier}},
  \bibinfo{author}{\bibfnamefont{P.}~\bibnamefont{Romatschke}},
  \bibinfo{author}{\bibfnamefont{D.~T.} \bibnamefont{Son}},
  \bibinfo{author}{\bibfnamefont{A.~O.} \bibnamefont{Starinets}},
  \bibnamefont{and} \bibinfo{author}{\bibfnamefont{M.~A.}
  \bibnamefont{Stephanov}}, \bibinfo{journal}{JHEP}
  \textbf{\bibinfo{volume}{04}}, \bibinfo{pages}{100} (\bibinfo{year}{2008}),
  \eprint{0712.2451}.

\bibitem[{\citenamefont{Denicol et~al.}(2012)\citenamefont{Denicol, Niemi,
  Molnar, and Rischke}}]{Denicol:2012cn}
\bibinfo{author}{\bibfnamefont{G.~S.} \bibnamefont{Denicol}},
  \bibinfo{author}{\bibfnamefont{H.}~\bibnamefont{Niemi}},
  \bibinfo{author}{\bibfnamefont{E.}~\bibnamefont{Molnar}}, \bibnamefont{and}
  \bibinfo{author}{\bibfnamefont{D.~H.} \bibnamefont{Rischke}},
  \bibinfo{journal}{Phys. Rev.} \textbf{\bibinfo{volume}{D85}},
  \bibinfo{pages}{114047} (\bibinfo{year}{2012}), \bibinfo{note}{[Erratum:
  Phys. Rev.D91,no.3,039902(2015)]}, \eprint{1202.4551}.

\bibitem[{\citenamefont{Finazzo et~al.}(2015)\citenamefont{Finazzo, Rougemont,
  Marrochio, and Noronha}}]{Finazzo:2014cna}
\bibinfo{author}{\bibfnamefont{S.~I.} \bibnamefont{Finazzo}},
  \bibinfo{author}{\bibfnamefont{R.}~\bibnamefont{Rougemont}},
  \bibinfo{author}{\bibfnamefont{H.}~\bibnamefont{Marrochio}},
  \bibnamefont{and} \bibinfo{author}{\bibfnamefont{J.}~\bibnamefont{Noronha}},
  \bibinfo{journal}{JHEP} \textbf{\bibinfo{volume}{02}}, \bibinfo{pages}{051}
  (\bibinfo{year}{2015}), \eprint{1412.2968}.

\bibitem[{\citenamefont{Cercignani and Kremer}(2002)}]{kremer}
\bibinfo{author}{\bibfnamefont{C.}~\bibnamefont{Cercignani}} \bibnamefont{and}
  \bibinfo{author}{\bibfnamefont{G.~M.} \bibnamefont{Kremer}},
  \emph{\bibinfo{title}{The Relativistic Boltzmann Equation: Theory and
  Applications}} (\bibinfo{publisher}{Birkhauser Verlag},
  \bibinfo{address}{Basel}, \bibinfo{year}{2002}).

\bibitem[{\citenamefont{Denicol
  et~al.}(2014{\natexlab{a}})\citenamefont{Denicol, Niemi, Bouras, Molnar, Xu,
  Rischke, and Greiner}}]{Denicol:2012vq}
\bibinfo{author}{\bibfnamefont{G.~S.} \bibnamefont{Denicol}},
  \bibinfo{author}{\bibfnamefont{H.}~\bibnamefont{Niemi}},
  \bibinfo{author}{\bibfnamefont{I.}~\bibnamefont{Bouras}},
  \bibinfo{author}{\bibfnamefont{E.}~\bibnamefont{Molnar}},
  \bibinfo{author}{\bibfnamefont{Z.}~\bibnamefont{Xu}},
  \bibinfo{author}{\bibfnamefont{D.~H.} \bibnamefont{Rischke}},
  \bibnamefont{and} \bibinfo{author}{\bibfnamefont{C.}~\bibnamefont{Greiner}},
  \bibinfo{journal}{Phys. Rev.} \textbf{\bibinfo{volume}{D89}},
  \bibinfo{pages}{074005} (\bibinfo{year}{2014}{\natexlab{a}}),
  \eprint{1207.6811}.

\bibitem[{\citenamefont{Denicol
  et~al.}(2014{\natexlab{b}})\citenamefont{Denicol, Heinz, Martinez, Noronha,
  and Strickland}}]{Denicol:2014xca}
\bibinfo{author}{\bibfnamefont{G.~S.} \bibnamefont{Denicol}},
  \bibinfo{author}{\bibfnamefont{U.~W.} \bibnamefont{Heinz}},
  \bibinfo{author}{\bibfnamefont{M.}~\bibnamefont{Martinez}},
  \bibinfo{author}{\bibfnamefont{J.}~\bibnamefont{Noronha}}, \bibnamefont{and}
  \bibinfo{author}{\bibfnamefont{M.}~\bibnamefont{Strickland}},
  \bibinfo{journal}{Phys. Rev. Lett.} \textbf{\bibinfo{volume}{113}},
  \bibinfo{pages}{202301} (\bibinfo{year}{2014}{\natexlab{b}}),
  \eprint{1408.5646}.

\bibitem[{\citenamefont{Denicol
  et~al.}(2014{\natexlab{c}})\citenamefont{Denicol, Heinz, Martinez, Noronha,
  and Strickland}}]{Denicol:2014tha}
\bibinfo{author}{\bibfnamefont{G.~S.} \bibnamefont{Denicol}},
  \bibinfo{author}{\bibfnamefont{U.~W.} \bibnamefont{Heinz}},
  \bibinfo{author}{\bibfnamefont{M.}~\bibnamefont{Martinez}},
  \bibinfo{author}{\bibfnamefont{J.}~\bibnamefont{Noronha}}, \bibnamefont{and}
  \bibinfo{author}{\bibfnamefont{M.}~\bibnamefont{Strickland}},
  \bibinfo{journal}{Phys. Rev.} \textbf{\bibinfo{volume}{D90}},
  \bibinfo{pages}{125026} (\bibinfo{year}{2014}{\natexlab{c}}),
  \eprint{1408.7048}.

\bibitem[{\citenamefont{Ryu et~al.}(2015)\citenamefont{Ryu, Paquet, Shen,
  Denicol, Schenke, Jeon, and Gale}}]{Ryu:2015vwa}
\bibinfo{author}{\bibfnamefont{S.}~\bibnamefont{Ryu}},
  \bibinfo{author}{\bibfnamefont{J.~F.} \bibnamefont{Paquet}},
  \bibinfo{author}{\bibfnamefont{C.}~\bibnamefont{Shen}},
  \bibinfo{author}{\bibfnamefont{G.~S.} \bibnamefont{Denicol}},
  \bibinfo{author}{\bibfnamefont{B.}~\bibnamefont{Schenke}},
  \bibinfo{author}{\bibfnamefont{S.}~\bibnamefont{Jeon}}, \bibnamefont{and}
  \bibinfo{author}{\bibfnamefont{C.}~\bibnamefont{Gale}},
  \bibinfo{journal}{Phys. Rev. Lett.} \textbf{\bibinfo{volume}{115}},
  \bibinfo{pages}{132301} (\bibinfo{year}{2015}), \eprint{1502.01675}.

\bibitem[{\citenamefont{Romatschke and Romatschke}(2017)}]{Romatschke:2017ejr}
\bibinfo{author}{\bibfnamefont{P.}~\bibnamefont{Romatschke}} \bibnamefont{and}
  \bibinfo{author}{\bibfnamefont{U.}~\bibnamefont{Romatschke}}
  (\bibinfo{year}{2017}), \eprint{1712.05815}.

\bibitem[{\citenamefont{Philipsen and Sch{\"a}fer}(2014)}]{Philipsen:2013nea}
\bibinfo{author}{\bibfnamefont{O.}~\bibnamefont{Philipsen}} \bibnamefont{and}
  \bibinfo{author}{\bibfnamefont{C.}~\bibnamefont{Sch{\"a}fer}},
  \bibinfo{journal}{JHEP} \textbf{\bibinfo{volume}{02}}, \bibinfo{pages}{003}
  (\bibinfo{year}{2014}).

\bibitem[{\citenamefont{Bemfica et~al.}(2018)\citenamefont{Bemfica, Disconzi,
  and Noronha}}]{Bemfica:2017wps}
\bibinfo{author}{\bibfnamefont{F.~S.} \bibnamefont{Bemfica}},
  \bibinfo{author}{\bibfnamefont{M.~M.} \bibnamefont{Disconzi}},
  \bibnamefont{and} \bibinfo{author}{\bibfnamefont{J.}~\bibnamefont{Noronha}},
  \bibinfo{journal}{Phys. Rev.} \textbf{\bibinfo{volume}{D98}},
  \bibinfo{pages}{104064} (\bibinfo{year}{2018}), \eprint{1708.06255}.

\bibitem[{\citenamefont{Disconzi et~al.}(2017)\citenamefont{Disconzi, Kephart,
  and Scherrer}}]{DisconziKephartScherrerNew}
\bibinfo{author}{\bibfnamefont{M.~M.} \bibnamefont{Disconzi}},
  \bibinfo{author}{\bibfnamefont{T.~W.} \bibnamefont{Kephart}},
  \bibnamefont{and} \bibinfo{author}{\bibfnamefont{R.~J.}
  \bibnamefont{Scherrer}}, \bibinfo{journal}{International Journal of Modern
  Physics D} \textbf{\bibinfo{volume}{26}}, \bibinfo{pages}{1750146 (52 pages)}
  (\bibinfo{year}{2017}).

\bibitem[{\citenamefont{Rezzolla and
  Zanotti}(2013)}]{RezzollaZanottiBookRelHydro}
\bibinfo{author}{\bibfnamefont{L.}~\bibnamefont{Rezzolla}} \bibnamefont{and}
  \bibinfo{author}{\bibfnamefont{O.}~\bibnamefont{Zanotti}},
  \emph{\bibinfo{title}{Relativistic Hydrodynamics}}
  (\bibinfo{publisher}{Oxford University Press}, \bibinfo{address}{New York},
  \bibinfo{year}{2013}).

\bibitem[{\citenamefont{Four\`es-Bruhat}(1958)}]{Choquet-BruhatFluidsExistence}
\bibinfo{author}{\bibfnamefont{Y.}~\bibnamefont{Four\`es-Bruhat}},
  \bibinfo{journal}{Bull. Soc. Math. France} \textbf{\bibinfo{volume}{86}},
  \bibinfo{pages}{155} (\bibinfo{year}{1958}), ISSN \bibinfo{issn}{0037-9484},
  \urlprefix\url{http://www.numdam.org/item?id=BSMF_1958__86__155_0}.

\bibitem[{\citenamefont{Choquet-Bruhat}(2009)}]{ChoquetBruhatGRBook}
\bibinfo{author}{\bibfnamefont{Y.}~\bibnamefont{Choquet-Bruhat}},
  \emph{\bibinfo{title}{General Relativity and the Einstein Equations}}
  (\bibinfo{publisher}{Oxford University Press}, \bibinfo{address}{New York},
  \bibinfo{year}{2009}).

\bibitem[{\citenamefont{Disconzi}(2015)}]{DisconziRemarksEinsteinEuler}
\bibinfo{author}{\bibfnamefont{M.~M.} \bibnamefont{Disconzi}},
  \bibinfo{journal}{Reviews in Mathematical Physics}
  \textbf{\bibinfo{volume}{27}}, \bibinfo{pages}{1550014}
  (\bibinfo{year}{2015}), \bibinfo{note}{45 pages}.

\bibitem[{\citenamefont{Lichnerowicz}(1967)}]{Lichnerowicz_MHD_book}
\bibinfo{author}{\bibfnamefont{A.}~\bibnamefont{Lichnerowicz}},
  \emph{\bibinfo{title}{Relativistic Hydrodynamics and Magnetohydrodynamics:
  Lectures on the Existence of Solutions}} (\bibinfo{publisher}{W. A.
  Benjamin}, \bibinfo{address}{New York}, \bibinfo{year}{1967}).

\bibitem[{\citenamefont{Hiscock and
  Lindblom}(1983)}]{Hiscock_Lindblom_stability_1983}
\bibinfo{author}{\bibfnamefont{W.~A.} \bibnamefont{Hiscock}} \bibnamefont{and}
  \bibinfo{author}{\bibfnamefont{L.}~\bibnamefont{Lindblom}},
  \bibinfo{journal}{Annals of Physics} \textbf{\bibinfo{volume}{151}},
  \bibinfo{pages}{466} (\bibinfo{year}{1983}).

\bibitem[{\citenamefont{Olson}(1990)}]{Olson:1989ey}
\bibinfo{author}{\bibfnamefont{T.~S.} \bibnamefont{Olson}},
  \bibinfo{journal}{Annals Phys.} \textbf{\bibinfo{volume}{199}},
  \bibinfo{pages}{18} (\bibinfo{year}{1990}).

\bibitem[{\citenamefont{Floerchinger and Grossi}(2018)}]{Floerchinger:2017cii}
\bibinfo{author}{\bibfnamefont{S.}~\bibnamefont{Floerchinger}}
  \bibnamefont{and} \bibinfo{author}{\bibfnamefont{E.}~\bibnamefont{Grossi}},
  \bibinfo{journal}{JHEP} \textbf{\bibinfo{volume}{08}}, \bibinfo{pages}{186}
  (\bibinfo{year}{2018}), \eprint{1711.06687}.

\bibitem[{\citenamefont{Courant and
  Hilbert}(1991)}]{Courant_and_Hilbert_book_2}
\bibinfo{author}{\bibfnamefont{C.}~\bibnamefont{Courant}} \bibnamefont{and}
  \bibinfo{author}{\bibfnamefont{D.}~\bibnamefont{Hilbert}},
  \emph{\bibinfo{title}{Methods of Mathematical Physics}},
  vol.~\bibinfo{volume}{2} (\bibinfo{publisher}{John Wiley \& Sons, Inc.},
  \bibinfo{year}{1991}), \bibinfo{edition}{1st} ed., ISBN
  \bibinfo{isbn}{0471504394}.

\bibitem[{\citenamefont{R.~P.~Geroch}(1996)}]{Geroch:1996kg}
\bibinfo{author}{\bibfnamefont{R.~P.} \bibnamefont{R.~P.~Geroch}}, in
  \emph{\bibinfo{booktitle}{General relativity. Proceedings, 46th Scottish
  Universities Summer School in Physics, NATO Advanced Study Institute,
  Aberdeen, UK, July 16-29, 1995}} (\bibinfo{year}{1996}),
  \eprint{gr-qc/9602055}.

\bibitem[{\citenamefont{Frittelli and Reula}(1996)}]{Frittelli:1996wr}
\bibinfo{author}{\bibfnamefont{S.}~\bibnamefont{Frittelli}} \bibnamefont{and}
  \bibinfo{author}{\bibfnamefont{O.~A.} \bibnamefont{Reula}},
  \bibinfo{journal}{Phys. Rev. Lett.} \textbf{\bibinfo{volume}{76}},
  \bibinfo{pages}{4667} (\bibinfo{year}{1996}), \eprint{gr-qc/9605005}.

\bibitem[{\citenamefont{Guermond et~al.}(2008)\citenamefont{Guermond, Marpeau,
  and Popov}}]{GuermondetalNumerical}
\bibinfo{author}{\bibfnamefont{J.-L.} \bibnamefont{Guermond}},
  \bibinfo{author}{\bibfnamefont{F.}~\bibnamefont{Marpeau}}, \bibnamefont{and}
  \bibinfo{author}{\bibfnamefont{B.}~\bibnamefont{Popov}},
  \bibinfo{journal}{Commun. Math. Sci.} \textbf{\bibinfo{volume}{6}},
  \bibinfo{pages}{199} (\bibinfo{year}{2008}), ISSN \bibinfo{issn}{1539-6746}.

\bibitem[{\citenamefont{Disconzi}(2014)}]{DisconziViscousFluidsNonlinearity}
\bibinfo{author}{\bibfnamefont{M.~M.} \bibnamefont{Disconzi}},
  \bibinfo{journal}{Nonlinearity} \textbf{\bibinfo{volume}{27}},
  \bibinfo{pages}{1915} (\bibinfo{year}{2014}).

\bibitem[{\citenamefont{Disconzi}(2019)}]{DisconziFollowupBemficaNoronha}
\bibinfo{author}{\bibfnamefont{M.~M.} \bibnamefont{Disconzi}},
  \bibinfo{journal}{Communications in Pure and Applied Analysis}
  \textbf{\bibinfo{volume}{18}}, \bibinfo{pages}{1567} (\bibinfo{year}{2019}).

\bibitem[{\citenamefont{Hawking and Ellis}(1975)}]{HawkingEllisBook}
\bibinfo{author}{\bibfnamefont{S.~W.} \bibnamefont{Hawking}} \bibnamefont{and}
  \bibinfo{author}{\bibfnamefont{G.~F.~R.} \bibnamefont{Ellis}},
  \emph{\bibinfo{title}{The Large Scale Structure of Space-Time (Cambridge
  Monographs on Mathematical Physics)}} (\bibinfo{publisher}{Cambridge
  University Press}, \bibinfo{year}{1975}), ISBN \bibinfo{isbn}{9780511524646},
  \urlprefix\url{https://doi.org/10.1017/CBO9780511524646}.

\bibitem[{\citenamefont{Wald}(2010)}]{WaldBookGR1984}
\bibinfo{author}{\bibfnamefont{R.~M.} \bibnamefont{Wald}},
  \emph{\bibinfo{title}{General relativity}} (\bibinfo{publisher}{University of
  Chicago press}, \bibinfo{year}{2010}).

\bibitem[{\citenamefont{Anile}(1990)}]{AnileBook}
\bibinfo{author}{\bibfnamefont{A.~M.} \bibnamefont{Anile}},
  \emph{\bibinfo{title}{Relativistic Fluids and Magneto-fluids: With
  Applications in Astrophysics and Plasma Physics (Cambridge Monographs on
  Mathematical Physics)}} (\bibinfo{publisher}{Cambridge University Press; 1
  edition}, \bibinfo{year}{1990}), ISBN \bibinfo{isbn}{9780511564130},
  \urlprefix\url{https://doi.org/10.1017/CBO9780511564130}.

\bibitem[{\citenamefont{Anile and Pennisi}(1987)}]{AnilePennisiMHD}
\bibinfo{author}{\bibfnamefont{A.~M.} \bibnamefont{Anile}} \bibnamefont{and}
  \bibinfo{author}{\bibfnamefont{S.}~\bibnamefont{Pennisi}},
  \bibinfo{journal}{Ann. Inst. H. Poincar\'e Phys. Th\'eor.}
  \textbf{\bibinfo{volume}{46}}, \bibinfo{pages}{27} (\bibinfo{year}{1987}),
  ISSN \bibinfo{issn}{0246-0211},
  \urlprefix\url{http://www.numdam.org/item?id=AIHPB_1987__46_1_27_0}.

\bibitem[{\citenamefont{Denicol et~al.}(2011)\citenamefont{Denicol, Noronha,
  Niemi, and Rischke}}]{Denicol:2011fa}
\bibinfo{author}{\bibfnamefont{G.~S.} \bibnamefont{Denicol}},
  \bibinfo{author}{\bibfnamefont{J.}~\bibnamefont{Noronha}},
  \bibinfo{author}{\bibfnamefont{H.}~\bibnamefont{Niemi}}, \bibnamefont{and}
  \bibinfo{author}{\bibfnamefont{D.~H.} \bibnamefont{Rischke}},
  \bibinfo{journal}{Phys. Rev.} \textbf{\bibinfo{volume}{D83}},
  \bibinfo{pages}{074019} (\bibinfo{year}{2011}), \eprint{1102.4780}.

\bibitem[{\citenamefont{Denicol et~al.}(2018)\citenamefont{Denicol, Gale, Jeon,
  Monnai, Schenke, and Shen}}]{Denicol:2018wdp}
\bibinfo{author}{\bibfnamefont{G.~S.} \bibnamefont{Denicol}},
  \bibinfo{author}{\bibfnamefont{C.}~\bibnamefont{Gale}},
  \bibinfo{author}{\bibfnamefont{S.}~\bibnamefont{Jeon}},
  \bibinfo{author}{\bibfnamefont{A.}~\bibnamefont{Monnai}},
  \bibinfo{author}{\bibfnamefont{B.}~\bibnamefont{Schenke}}, \bibnamefont{and}
  \bibinfo{author}{\bibfnamefont{C.}~\bibnamefont{Shen}},
  \bibinfo{journal}{Phys. Rev.} \textbf{\bibinfo{volume}{C98}},
  \bibinfo{pages}{034916} (\bibinfo{year}{2018}), \eprint{1804.10557}.

\bibitem[{\citenamefont{Jeon and Heinz}(2015)}]{Jeon:2015dfa}
\bibinfo{author}{\bibfnamefont{S.}~\bibnamefont{Jeon}} \bibnamefont{and}
  \bibinfo{author}{\bibfnamefont{U.}~\bibnamefont{Heinz}},
  \bibinfo{journal}{Int. J. Mod. Phys.} \textbf{\bibinfo{volume}{E24}},
  \bibinfo{pages}{1530010} (\bibinfo{year}{2015}), \eprint{1503.03931}.

\bibitem[{\citenamefont{Eckart}(1940)}]{EckartViscous}
\bibinfo{author}{\bibfnamefont{C.}~\bibnamefont{Eckart}},
  \bibinfo{journal}{Physical Review} \textbf{\bibinfo{volume}{58}},
  \bibinfo{pages}{919} (\bibinfo{year}{1940}).

\bibitem[{\citenamefont{Christodoulou}(2007)}]{ChristodoulouShocks}
\bibinfo{author}{\bibfnamefont{D.}~\bibnamefont{Christodoulou}},
  \emph{\bibinfo{title}{The formation of shocks in 3-dimensional fluids}}, EMS
  Monographs in Mathematics (\bibinfo{publisher}{European Mathematical Society
  (EMS), Z\"urich}, \bibinfo{year}{2007}), ISBN
  \bibinfo{isbn}{978-3-03719-031-9},
  \urlprefix\url{https://doi.org/10.4171/031}.

\bibitem[{\citenamefont{Speck}(2012)}]{SpeckNonlinearStability}
\bibinfo{author}{\bibfnamefont{J.}~\bibnamefont{Speck}},
  \bibinfo{journal}{Selecta Mathematica} \textbf{\bibinfo{volume}{18}},
  \bibinfo{pages}{633} (\bibinfo{year}{2012}).

\bibitem[{\citenamefont{Rendall}(1992)}]{RendallFluidBodies}
\bibinfo{author}{\bibfnamefont{A.~D.} \bibnamefont{Rendall}},
  \bibinfo{journal}{J. Math. Phys.} \textbf{\bibinfo{volume}{33}},
  \bibinfo{pages}{1047} (\bibinfo{year}{1992}), ISSN \bibinfo{issn}{0022-2488},
  \urlprefix\url{http://dx.doi.org/10.1063/1.529766}.

\bibitem[{\citenamefont{Heinz and Snellings}(2013)}]{Heinz:2013th}
\bibinfo{author}{\bibfnamefont{U.}~\bibnamefont{Heinz}} \bibnamefont{and}
  \bibinfo{author}{\bibfnamefont{R.}~\bibnamefont{Snellings}},
  \bibinfo{journal}{Ann. Rev. Nucl. Part. Sci.} \textbf{\bibinfo{volume}{63}},
  \bibinfo{pages}{123} (\bibinfo{year}{2013}), \eprint{1301.2826}.

\bibitem[{\citenamefont{Noronha-Hostler
  et~al.}(2013)\citenamefont{Noronha-Hostler, Denicol, Noronha, Andrade, and
  Grassi}}]{Noronha-Hostler:2013gga}
\bibinfo{author}{\bibfnamefont{J.}~\bibnamefont{Noronha-Hostler}},
  \bibinfo{author}{\bibfnamefont{G.~S.} \bibnamefont{Denicol}},
  \bibinfo{author}{\bibfnamefont{J.}~\bibnamefont{Noronha}},
  \bibinfo{author}{\bibfnamefont{R.~P.~G.} \bibnamefont{Andrade}},
  \bibnamefont{and} \bibinfo{author}{\bibfnamefont{F.}~\bibnamefont{Grassi}},
  \bibinfo{journal}{Phys. Rev.} \textbf{\bibinfo{volume}{C88}},
  \bibinfo{pages}{044916} (\bibinfo{year}{2013}), \eprint{1305.1981}.

\bibitem[{\citenamefont{Noronha-Hostler
  et~al.}(2014)\citenamefont{Noronha-Hostler, Noronha, and
  Grassi}}]{Noronha-Hostler:2014dqa}
\bibinfo{author}{\bibfnamefont{J.}~\bibnamefont{Noronha-Hostler}},
  \bibinfo{author}{\bibfnamefont{J.}~\bibnamefont{Noronha}}, \bibnamefont{and}
  \bibinfo{author}{\bibfnamefont{F.}~\bibnamefont{Grassi}},
  \bibinfo{journal}{Phys. Rev.} \textbf{\bibinfo{volume}{C90}},
  \bibinfo{pages}{034907} (\bibinfo{year}{2014}), \eprint{1406.3333}.

\bibitem[{\citenamefont{Torrieri et~al.}(2008)\citenamefont{Torrieri, Tomasik,
  and Mishustin}}]{Torrieri:2007fb}
\bibinfo{author}{\bibfnamefont{G.}~\bibnamefont{Torrieri}},
  \bibinfo{author}{\bibfnamefont{B.}~\bibnamefont{Tomasik}}, \bibnamefont{and}
  \bibinfo{author}{\bibfnamefont{I.}~\bibnamefont{Mishustin}},
  \bibinfo{journal}{Phys. Rev.} \textbf{\bibinfo{volume}{C77}},
  \bibinfo{pages}{034903} (\bibinfo{year}{2008}), \eprint{0707.4405}.

\bibitem[{\citenamefont{Rajagopal and Tripuraneni}(2010)}]{Rajagopal:2009yw}
\bibinfo{author}{\bibfnamefont{K.}~\bibnamefont{Rajagopal}} \bibnamefont{and}
  \bibinfo{author}{\bibfnamefont{N.}~\bibnamefont{Tripuraneni}},
  \bibinfo{journal}{JHEP} \textbf{\bibinfo{volume}{03}}, \bibinfo{pages}{018}
  (\bibinfo{year}{2010}), \eprint{0908.1785}.

\bibitem[{\citenamefont{Denicol et~al.}(2015)\citenamefont{Denicol, Gale, and
  Jeon}}]{Denicol:2015bpa}
\bibinfo{author}{\bibfnamefont{G.~S.} \bibnamefont{Denicol}},
  \bibinfo{author}{\bibfnamefont{C.}~\bibnamefont{Gale}}, \bibnamefont{and}
  \bibinfo{author}{\bibfnamefont{S.}~\bibnamefont{Jeon}},
  \bibinfo{journal}{PoS} \textbf{\bibinfo{volume}{CPOD2014}},
  \bibinfo{pages}{033} (\bibinfo{year}{2015}), \eprint{1503.00531}.

\bibitem[{\citenamefont{Maartens}(1995)}]{MaartensDissipative}
\bibinfo{author}{\bibfnamefont{R.}~\bibnamefont{Maartens}},
  \bibinfo{journal}{Class. Quantum Grav.} \textbf{\bibinfo{volume}{12}},
  \bibinfo{pages}{1455} (\bibinfo{year}{1995}).

\bibitem[{\citenamefont{Li and Barrow}(2009)}]{LiBarrow}
\bibinfo{author}{\bibfnamefont{B.}~\bibnamefont{Li}} \bibnamefont{and}
  \bibinfo{author}{\bibfnamefont{J.~D.} \bibnamefont{Barrow}},
  \bibinfo{journal}{Phys. Rev. D} \textbf{\bibinfo{volume}{79}},
  \bibinfo{pages}{103521} (\bibinfo{year}{2009}).

\bibitem[{\citenamefont{Disconzi et~al.}(2015)\citenamefont{Disconzi, Kephart,
  and Scherrer}}]{Disconzi_Kephart_Scherrer_2015}
\bibinfo{author}{\bibfnamefont{M.~M.} \bibnamefont{Disconzi}},
  \bibinfo{author}{\bibfnamefont{T.~W.} \bibnamefont{Kephart}},
  \bibnamefont{and} \bibinfo{author}{\bibfnamefont{R.~J.}
  \bibnamefont{Scherrer}}, \bibinfo{journal}{Physical Review D}
  \textbf{\bibinfo{volume}{91}}, \bibinfo{pages}{043532 (6 pages)}
  (\bibinfo{year}{2015}).

\bibitem[{\citenamefont{Zimdahl et~al.}(2001)\citenamefont{Zimdahl, Schwarz,
  Balakin, and Pavon}}]{Zimdahl}
\bibinfo{author}{\bibfnamefont{W.}~\bibnamefont{Zimdahl}},
  \bibinfo{author}{\bibfnamefont{D.}~\bibnamefont{Schwarz}},
  \bibinfo{author}{\bibfnamefont{A.~B.} \bibnamefont{Balakin}},
  \bibnamefont{and} \bibinfo{author}{\bibfnamefont{D.}~\bibnamefont{Pavon}},
  \bibinfo{journal}{Phys. Rev. D} \textbf{\bibinfo{volume}{64}},
  \bibinfo{pages}{063501} (\bibinfo{year}{2001}).

\bibitem[{\citenamefont{Piattella et~al.}(2011)\citenamefont{Piattella, Fabris,
  and Zimdahl}}]{Piattella_et_al}
\bibinfo{author}{\bibfnamefont{O.~F.} \bibnamefont{Piattella}},
  \bibinfo{author}{\bibfnamefont{J.~C.} \bibnamefont{Fabris}},
  \bibnamefont{and} \bibinfo{author}{\bibfnamefont{W.}~\bibnamefont{Zimdahl}},
  \bibinfo{journal}{JCAP} \textbf{\bibinfo{volume}{1105}}, \bibinfo{pages}{029}
  (\bibinfo{year}{2011}).

\bibitem[{\citenamefont{Montani and
  Venanzi}(2017)}]{MontaniLichnerowiczViscosityBianchi}
\bibinfo{author}{\bibfnamefont{G.}~\bibnamefont{Montani}} \bibnamefont{and}
  \bibinfo{author}{\bibfnamefont{M.}~\bibnamefont{Venanzi}},
  \bibinfo{journal}{Eur. Phys. J.} \textbf{\bibinfo{volume}{C77}},
  \bibinfo{pages}{486} (\bibinfo{year}{2017}).

\bibitem[{\citenamefont{East et~al.}(2016)\citenamefont{East, Kleban, Linde,
  and Senatore}}]{East:2015ggf}
\bibinfo{author}{\bibfnamefont{W.~E.} \bibnamefont{East}},
  \bibinfo{author}{\bibfnamefont{M.}~\bibnamefont{Kleban}},
  \bibinfo{author}{\bibfnamefont{A.}~\bibnamefont{Linde}}, \bibnamefont{and}
  \bibinfo{author}{\bibfnamefont{L.}~\bibnamefont{Senatore}},
  \bibinfo{journal}{JCAP} \textbf{\bibinfo{volume}{1609}}, \bibinfo{pages}{010}
  (\bibinfo{year}{2016}), \eprint{1511.05143}.

\bibitem[{\citenamefont{Eyink and Drivas}(2018)}]{Eyink:2017zfz}
\bibinfo{author}{\bibfnamefont{G.~L.} \bibnamefont{Eyink}} \bibnamefont{and}
  \bibinfo{author}{\bibfnamefont{T.~D.} \bibnamefont{Drivas}},
  \bibinfo{journal}{Phys. Rev.} \textbf{\bibinfo{volume}{X8}},
  \bibinfo{pages}{011023} (\bibinfo{year}{2018}), \eprint{1704.03541}.

\bibitem[{\citenamefont{Disconzi and Speck}(2018)}]{DisconziSpeckRelEulerNull}
\bibinfo{author}{\bibfnamefont{M.~M.} \bibnamefont{Disconzi}} \bibnamefont{and}
  \bibinfo{author}{\bibfnamefont{J.}~\bibnamefont{Speck}},
  \bibinfo{journal}{arXiv:1809.06204 [math.AP]}  (\bibinfo{year}{2018}).

\bibitem[{\citenamefont{Kato}(1975)}]{KatoQuasiLinear}
\bibinfo{author}{\bibfnamefont{T.}~\bibnamefont{Kato}}, \bibinfo{journal}{Arch.
  Rational Mech. Anal} \textbf{\bibinfo{volume}{58}}, \bibinfo{pages}{181}
  (\bibinfo{year}{1975}).

\bibitem[{\citenamefont{Fischer and
  Marsden}(1972)}]{Fischer_Marsden_Einstein_FOSH_1972}
\bibinfo{author}{\bibfnamefont{A.~E.} \bibnamefont{Fischer}} \bibnamefont{and}
  \bibinfo{author}{\bibfnamefont{J.~E.} \bibnamefont{Marsden}},
  \bibinfo{journal}{Commun. Math. Phys.} \textbf{\bibinfo{volume}{28}},
  \bibinfo{pages}{1} (\bibinfo{year}{1972}).

\bibitem[{\citenamefont{Evans}(2010)}]{EvansPDE}
\bibinfo{author}{\bibfnamefont{L.~C.} \bibnamefont{Evans}},
  \emph{\bibinfo{title}{Partial differential equations}},
  vol.~\bibinfo{volume}{19} of \emph{\bibinfo{series}{Graduate Studies in
  Mathematics}} (\bibinfo{publisher}{American Mathematical Society, Providence,
  RI}, \bibinfo{year}{2010}), \bibinfo{edition}{2nd} ed., ISBN
  \bibinfo{isbn}{978-0-8218-4974-3}.

\bibitem[{\citenamefont{Majda}(1984)}]{MajdaCompressibleFlow}
\bibinfo{author}{\bibfnamefont{A.}~\bibnamefont{Majda}},
  \emph{\bibinfo{title}{Compressible fluid flow and systems of conservation
  laws in several space variables}}, vol.~\bibinfo{volume}{53} of
  \emph{\bibinfo{series}{Applied Mathematical Sciences}}
  (\bibinfo{publisher}{Springer-Verlag, New York}, \bibinfo{year}{1984}), ISBN
  \bibinfo{isbn}{0-387-96037-6},
  \urlprefix\url{http://dx.doi.org/10.1007/978-1-4612-1116-7}.

\end{thebibliography}

\appendix

\section{Appendix}

\section{Causality}

Here we give a precise formulation of the causality of the generalized EIS system and its proof.

\begin{theorem}
Let $(\ep,u^\alpha, \Pi, n, g_{\alpha \beta})$ be a solution to the generalized EIS equations defined
on a globally hyperbolic spacetime $M$, and let $\Sigma$ be a Cauchy surface. Suppose $\frac{\zeta}{\tau_\Pi(\varepsilon+P+\Pi)}+\alpha_1 
+ \frac{\alpha_2 \,n}{\varepsilon + P+\Pi} \geq 0$ and that (1) holds. 
Then, for any $p \in M$ in the future of $\Sigma$, 
$(\ep(p),u^\alpha(p), \Pi (p), n(p), g_{\alpha \beta}(p))$ depends only 
on $\left. (\ep,u^\alpha, \Pi, n, g_{\alpha \beta}, \kappa_{\alpha\beta})\right|_{\Sigma \cap J^-(p)}$,
where $J^-(p)$ is the causal past of $p$ and $\kappa$ is the extrinsic curvature of $\Sigma$
in $M$.
\label{theorem_causality}
\end{theorem}

Theorem \ref{theorem_causality} gives a precise mathematical formulation of the statement 
that in the generalized EIS system faster-than-light signals are absent and the future
cannot influence the past. The concept of causality is properly
formulated in globally hyperbolic spacetimes since even solutions to vacuum Einstein's equations can
otherwise fail to be causal \cite{HawkingEllisBook,WaldBookGR1984}. 
A Cauchy surface plays the role of the set $\{ x^0 = 
\text{constant} \}$ when coordinates are introduced, with $x^0$ playing the role of time.
However,  Theorem \ref{theorem_causality} is formulated with no mention to coordinates
because a precise statement of causality in general relativity has to be coordinate free.
The causal past of $p$, $J^-(p)$, is the general relativistic analogue 
of the past light-cone at a point \cite{HawkingEllisBook,WaldBookGR1984}. The conclusion of the theorem says that the values 
of the fields $\ep,u_\alpha, \Pi$, and $g_{\alpha \beta}$ at $p$ are determined
only by the the dynamics of such fields in the (causal) past of $p$. 
The curvature $\kappa_{\alpha\beta}$
can be thought of as the normal derivative
of $g_{\alpha\beta}$ along $\Sigma$. 
To understand the role of $\kappa_{\alpha\beta}$ note that, in
the most common situation, $\Sigma$ is a surface
where initial data for the generalized EIS system is provided. In this case
the metric and its first derivatives have to be prescribed on $\Sigma$ since
Einstein's equations are of second order. 

\noindent \emph{Proof of Theorem 1:}
To prove Theorem 
1, we need to compute the characteristics of the generalized EIS 
system \cite{Courant_and_Hilbert_book_2}.
First, we write equations 
(2), (3), (4), and (5) in matrix form as
$\cM^\mu \nabla_\mu \Psi + \cN \Psi = 0$, where $\Psi = (\varepsilon, u^\nu, n, \Pi)$ and the matrices
$\cM^\mu$ are given by
\bea
\begin{pmatrix}
  u^\mu \qquad & \left(\varepsilon+P+\Pi\right) \delta^\mu_\nu & 0 & 0 \\
  \alpha_1\Delta^{\alpha\mu} \qquad & 1_{4\times 4}\left(\varepsilon+P+\Pi \right)u^\mu  & \alpha_2 \Delta^{\alpha\mu} & 
  \Delta^{\alpha\mu} \\
  0 & n \delta^\mu_\nu & u^\mu & 0\\
    0 & \zeta \delta^\mu_\nu & 0 & \tau_\Pi u^\mu  
 \end{pmatrix},
\nonumber
\eea
and $\cN$ is given by
\bea
\begin{pmatrix}
  0 & 0_{1\times 4} & 0 & 0  \\
  0_{4\times 1} & 0_{4\times 4} & 0_{4\times 1}  & 0_{4\times 1}\\
   0& 0_{1\times4}& 0 & 0\\
     0 & 0_{1\times 4} & 0 & 1 + \lambda \Pi
 \end{pmatrix}.
\nonumber
\eea
Fix a point $p$ in spacetime. Choose coordinates
$\{ x^\mu \}$ near $p$. 
Expanding the covariant derivative, the system reads
$\cM^\mu \partial_\mu \Psi +\cM_\Gamma\Psi + \cN \Psi= 0$, where $\cM_\Gamma$ denotes
the matrix with terms involving the Christoffel symbols coming from expanding $\nabla_\mu$.

Letting $\xi_\mu$ be an arbitrary covector in the cotangent space at $p$, the characteristic 
determinant of the system is $\cP(\xi) =\det( \cM^\mu \xi_\mu )$. Computing, we find
\begin{align}
\mathcal{P}(\xi)& =
 (u^\alpha \xi_\alpha)^5\left(\varepsilon+P+\Pi \right)^4\tau_\Pi\Big[(u^\alpha \xi_\alpha)^2 
 \nonumber
 \\
 & \quad - \Big(\frac{\zeta}{\tau_\Pi(\varepsilon+P+\Pi)}+\alpha_1 
 \nonumber 
 \\
 &\quad + \frac{\alpha_2 \,n}{\varepsilon + P+\Pi}\Big)\Delta^{\mu\lambda}\xi_\mu\xi_\lambda\Big].  
  \label{char_det}
\end{align}
The system's characteristics are determined by the $\xi_\mu$'s such that $P(\xi) = 0$.
Assuming condition 
(1), we have $\varepsilon+P+\Pi \neq 0$, $\tau_\Pi \neq 0$, and the characteristics
are thus determined by $u^\alpha \xi_\alpha = 0$ or by the solutions coming from setting the term in square
brackets in \eqref{char_det} to zero. The characteristics corresponding to 
$u^\alpha \xi_\alpha = 0$ are simply the flow lines of $u^\alpha$, which are causal as long as $u^\alpha$ remains timelike. The characteristics associated with the vanishing of the bracket in \eqref{char_det}
are precisely the characteristics of an acoustical metric \cite{ChristodoulouShocks,DisconziSpeckRelEulerNull} with an effective sound speed squared given by
$c_{\text{acoustical}}^2 = \frac{\zeta}{\tau_\Pi(\varepsilon+P+\Pi)}+\alpha_1 
+ \frac{\alpha_2 \,n}{\varepsilon + P+\Pi}\geq 0$. Thus, the system will be causal as long as
$u^\alpha$ remains timelike (and normalized) and 
$c_{\text{acoustical}}^2 \leq 1$, which is precisely condition (1). 

The discussion above has not taken into account Einstein's equations (6). 
We consider now the full system 
(2), (3), (4), (5), and (6), and assume that our coordinates about $p$ are 
wave coordinates \cite{ChoquetBruhatGRBook}, in which case Einstein's equations can be written as
\begin{align}
-\frac{1}{2} g^{\mu\nu} \partial_\mu \partial_\nu g_{\alpha\beta} &= 
8\pi G T_{\alpha\beta} -\frac{1}{2}8\pi G  g^{\mu\nu} T_{\mu\nu} g_{\alpha\beta} 
\nonumber
\\
& \quad + \Lambda g_{\alpha\beta}
+ \cZ_{\alpha\beta},
\label{EE_wave}
\end{align}
where $\cZ_{\alpha \beta}$ involves only $g_{\alpha\beta}$ and its first derivatives. Since
$T_{\alpha\beta}$ involves no derivatives of $\Psi$ and equations 
(2), (3), (4), (5), involve at most one derivative of
$g_{\alpha\beta}$ (via the Christoffel symbols that have been grouped in the matrix $M_\Gamma$),
the characteristic determinant of the system is now $\cP(\xi) \cQ(\xi)$, where $\cP(\xi)$
is as above and $\cQ(\xi) = (g^{\mu\nu}\xi_\mu \xi_\nu)^{10}$. 
$\cQ(\xi)=0$ corresponds to the characteristics associated with the light-cones, hence causal.
Standard arguments from general relativity show that the characteristics computed in this
fashion are geometric or coordinate independent \cite{WaldBookGR1984}. \hfill $\cqd$

\section{Proof of Theorem 2}

Denote by $H_{\text{ul}}^s$ the uniformly local Sobolev space (functions whose distributional derivatives up to order
$s$ exist and are square-integrable in a partition), 
and by $C^s$ the space of continuously diffentiable functions up to order $s$ \cite{ChoquetBruhatGRBook,KatoQuasiLinear}. We have:

\begin{theorem}
Let $\cI = (\Sigma, \mathring{\ep},\mathring{u}^i, \mathring{\Pi}, \mathring{n},
\mathring{g}_{ij}, \mathring{\kappa}_{ij})$ be an initial data set
for the generalized EIS system, with an equation of state $P = P(\varepsilon, n)$,
a bulk viscosity $\zeta = \zeta(\varepsilon,n)$, and a relaxation time $\tau_\Pi=\tau_\Pi(\varepsilon,n)$. Assume that $\mathring{\varepsilon}+P(\mathring{\varepsilon}, \mathring{n})+\mathring{\Pi}$, $\tau_\Pi (\mathring{\varepsilon}, \mathring{n})$, $\zeta(\mathring{\varepsilon}, \mathring{n})>0$, $\frac{\partial P}{\partial \varepsilon}(\mathring{\varepsilon},\mathring{n}) + \frac{\partial P}{\partial n}(\mathring{\varepsilon},\mathring{n}) \mathring{n}/(\mathring{\varepsilon}+P(\mathring{\varepsilon}, \mathring{n})+\mathring{\Pi})\ge 0$, 
and that   
$\mathring{n}$, $\frac{\partial P}{\partial \varepsilon}(\mathring{\varepsilon},\mathring{n})$, and $\frac{\partial P}{\partial n}(\mathring{\varepsilon},\mathring{n})$ are nonzero.
Suppose that
$\mathring{g}_{ij} \in H^{s+1}_{\text{ul}}(\Sigma)$, 
$\mathring{\ep},\mathring{u}^i, \mathring{\Pi}, \mathring{n},
\mathring{\kappa}_{ij} \in H^s_{\text{ul}}(\Sigma)$,  and that
$P,\zeta, \lambda, \tau_\Pi \in C^s(\mathbb{R}^2)$, where $s\geq 3$.
Suppose that (1) 
holds for $\cI$. Then, there exists a globally
hyperbolic development of $\cI$. 
\label{theorem_symmetric_hyperbolic}
\end{theorem}
An initial data set for the generalized EIS system involves a three-dimensional
manifold $\Sigma$ that plays the analogue of a slice of ``constant time" (this is just an
analogy since constant time is a coordinate dependent notion). The initial data 
 along $\Sigma$ cannot be arbitrary but need to satisfy the 
constraint equations \cite{HawkingEllisBook,WaldBookGR1984}. 

Since a solution to Einstein's equations is not a set of fields on spacetime but spacetime
itself, existence and uniqueness of solutions is properly stated in the context of general relativity
as the existence of a globally hyperbolic development. Also, for Einstein's equations,
uniqueness is meant in a geometric sense \cite{HawkingEllisBook,WaldBookGR1984}.

We begin showing that equations (2), (3), (4), and (5) 
can be written as a FOSH system.
We use that $\nabla_\alpha u^0 = -\frac{u_i}{u_0}\nabla_\alpha u^i$, multiply (3) 
by $C_i^\alpha =  g_i^\alpha - \frac{u_i}{u_0}\delta^\alpha_0$,
and use that $C_j^\alpha u_\alpha=0$.
Assuming that  $\zeta, \tau_\Pi, (\varepsilon+P+\Pi)>0$ and $n, \left(\frac{\partial P}{\partial \varepsilon}\right)_n,\left(\frac{\partial P}{\partial n}\right)_\varepsilon \neq 0$, we readily see that 
equations  (2), (3), (4), and (5) 
assume the form $\cA^\mu\nabla_\mu \Phi + \cC \Phi = 0$, 
where the
matrices $\cA^\mu$ and $\Phi$ are as in Section 2.2, and $\cC$ is the matrix that has all entries equal to zero except for
$\cC_{6,6}= \frac{1+\lambda \Pi}{\zeta \left(\frac{\partial P}{\partial \varepsilon}\right)_n}$.
Expanding the covariant derivative we have
$\cA^\mu\partial_\mu \Phi + \cA_\Gamma \Phi + \cC \Phi = 0$, where
$\cA_\Gamma$ denotes the matrix with terms involving Christoffel symbols. Thus, we have
$\cA^\mu\partial_\mu \Phi + \cB = 0$, with $\cB = \cA_\Gamma \Phi + \cC \Phi$.
Note that $\cB$ involves no derivatives of $\Phi$. Below, we show that $\cA^0$ is positive 
definite.

Embed $\Sigma$ into $\RR \times \Sigma$ and fix a point $p \in \Sigma$. We consider
Einstein's equations written in wave coordinates in a neighborhood of $p$, i.e., 
equation \eqref{EE_wave}. Without loss of generality we can assume that the coordinates
are such that $\{x^i\}$ are coordinates on $\Sigma$ and that $g_{\alpha\beta}$ is the Minkowski
metric at $p$ \cite{WaldBookGR1984}.
It has long been known that Einstein's equations can be written as 
a FOSH system \cite{Fischer_Marsden_Einstein_FOSH_1972}. Indeed,
setting $\Theta = (g_{\alpha\beta}, \partial_i g_{\alpha\beta}, \partial_0 g_{\alpha\beta})$,
\eqref{EE_wave} reads  $\widetilde{A}^\mu \partial_\mu \Theta + \widetilde{\cB} = 0$,
where  $\widetilde{\cA}^0$ and $\widetilde{\cA}^k$ are given by, respectively,
\bea
\begin{pmatrix}
1_{10} & 0_{10} & 0_{10} & 0_{10} & 0_{10} \\
0_{10} & g^{11} 1_{10} & g^{12} 1_{10} & g^{13} 1_{10} & 0_{10} \\
0_{10} & g^{12} 1_{10} & g^{22} 1_{10} & g^{23} 1_{10} & 0_{10} \\
0_{10} & g^{13} 1_{10} & g^{23} 1_{10} & g^{33} 1_{10} & 0_{10} \\
0_{10} & 0_{10} & 0_{10} & 0_{10} & -g^{00} 1_{10} \\
\end{pmatrix},
\nonumber
\eea
and
\bea
\begin{pmatrix}
0_{10} & 0_{10} & 0_{10} & 0_{10} & 0_{10} \\
0_{10} & 0_{10} & 0_{10} & 0_{10} & g^{k1} 1_{10} \\
0_{10} & 0_{10} & 0_{10} & 0_{10} & g^{k2} 1_{10} \\
0_{10} & 0_{10} & 0_{10} & 0_{10} & g^{k3} 1_{10} \\
0_{10} & g^{k1} 1_{10\times 10} & g^{k2} 1_{10} & g^{k3} 1_{10} & 
2 g^{k0} 1_{10} 
\end{pmatrix},
\nonumber
\eea
where $1_{10}$ and $0_{10}$ are the $10\times 10$ identity and zero matrices, respectively,
and $\widetilde{\cB}$ contains the terms on the right-hand side of \eqref{EE_wave}.
The full system thus reads
$\mathfrak{A}^\mu \partial_\mu \Xi + \mathfrak{B} = 0$, where
$\Xi = (\Phi,\Theta)$, $\mathfrak{B} = (\cB, \widetilde{\cB})$, and
\bea
\mathfrak{A}^\mu = 
\begin{pmatrix}
\cA^\mu & 0 \\
0 & \widetilde{\cA}^\mu 
\end{pmatrix}.
\nonumber
\eea
The matrices $\mathfrak{A}^\mu$ are symmetric and $\mathfrak{B}$ does not involve any 
derivative of $\Xi$. In $\mathfrak{B}$ we have already eliminated $u^0$ appearing 
on the right-hand side of \eqref{EE_wave} in terms of $u^i$, so that $\mathfrak{B}$
depends only on $\Xi$. To see that such elimination is indeed possible, note that
at $p$ the normalization $u^\mu u_\mu = -1$ reads
$(u^0)^2 = 1+ \delta_{ij} u^i u^j$, and we define $u^0$ as the positive solution
(since $u^\mu$ is future directed). Below we argue that the fields vary continuously with
the spacetime coordinates. Thus, by continuity, the equation $u^\mu u_\mu = -1$ has precisely
two solutions $u^0$ in a neighborhood of $p$, one, and only one, of which determines
a future pointing vector.

Since $\widetilde{\cA}^0$ is known to be positive definite
\cite{Fischer_Marsden_Einstein_FOSH_1972}, to conclude that $\mathfrak{A}^0$
is positive definite it suffices to analyze $\cA^0$. We do this by verifying that all
the leading principal minors of $\cA^0$ are positive. The $5\times 5$ upper left part of $\cA^0$ has all minors positive, since it
corresponds to the ideal fluid which is already known to form a symmetric hyperbolic
system for $(\varepsilon, u^i, n)$. 

It remains to compute the determinant of $\cA^0$.
For this, we first compute $\det \cA^0$ at $p$, in which case
$u_\mu = 1/\sqrt{1-v^2}(-1,\mathbf{v})$ (recall that $g_{\alpha\beta}$ is the Minkowski metric
at $p$). 
Then
\begin{align}
\det \mathcal{A}^0 &=  \frac{(\varepsilon+P
 +\Pi)}{n \alpha_1^5}\alpha_2 u_0^4 \Big[(\varepsilon+P+\Pi)\frac{\tau_\Pi}{\zeta}(1
 \nonumber
 \\
 & \quad -K v^2) -v^2\Big],
\nonumber
\end{align}
where $v^2 = \mathbf{v}^2$,with $v \in [0,1)$, and
$K = \alpha_1 + \frac{\alpha_2 \,n}{\varepsilon+P+\Pi}$.
Since 
\begin{align}
(\varepsilon+P+\Pi)\frac{\tau_\Pi}{\zeta}\left(1-K v^2\right) -v^2 
\nonumber \\
> (\varepsilon+P+\Pi)\frac{\tau_\Pi}{\zeta}\left(1-K\right) -1
\nonumber
\end{align}
if $K\ge 0$, we see that $\det \mathcal{A}^0>0$ if 
\be
(\varepsilon+P+\Pi)\frac{\tau_\Pi}{\zeta}\left(1-K\right) -1 \geq  0,
\nonumber
\ee
but this later condition is the same as (1). 
Because the fields are
continuous functions of the spacetime coordinates, we conclude that 
$\det \mathcal{A}^0>0$ will also hold in a neighborhood of $p$.

In order to invoke theorems of hyperbolic differential equations to establish
Theorem 2, 
we need to check that the system
is  FOSH when the fields take the initial data (the solution $\Xi$ 
has yet to be shown to exist; the only fields at our disposal at this point
are coming from the initial data). But this is immediate since  
$\mathfrak{A}$ and $\mathfrak{B}$ do not depend on derivatives of $\Xi$ and the foregoing conditions hold for the initial data by hypothesis. Furthermore, the continuity
required for the foregoing arguments is true because functions in $H^s$ are in particular
continuous when $s\geq 3$ in view of the standard Sobolev embedding theorem \cite{EvansPDE}. By
standard existence and uniqueness theorems for FOSH systems \cite{ChoquetBruhatGRBook,MajdaCompressibleFlow} we obtain a solution in a neighborhood
of $p$. It is known that a solution to Einstein's equations in wave coordinates
yields a solution to the full Einstein equations if and only if the constraints are satisfied, which 
is the case by hypothesis. Moreover, defining $u^0$ via the relation $u^\mu u_\mu = -1$
(which uniquely defines $u^0$ in a neighborhood of $p$, as above) and working backwards,
we see that the original equations (2), (3), (4), (5), and (6) 
are satisfied in a neighborhood of $p$.
Finally, a solution valid in a neighborhood of $\Sigma$ is obtained
by a standard argument of constructing solutions as above in the neighborhood of different
points $p$ and gluing them via diffeomorphisms \cite{WaldBookGR1984}. Standard properties of Einstein's equations
give the existence of the desired globally hyperbolic development \cite{ChoquetBruhatGRBook}.

\end{document}